\documentclass[11pt,a4paper]{article}
\pdfoutput=1 
\usepackage{jinstpub}
\usepackage{siunitx}
\usepackage{adjustbox}
\usepackage{subfigure}

\title{First measurement of nuclear recoil head-tail sense in a fiducialised WIMP dark matter detector \boldmath}
\collaboration{DRIFT Collaboration:}
\author[a]{J.~B.~R.~Battat,} 

\author[b]{E.~Daw,}
\author[b,1]{A.~C.~Ezeribe, }
\note{Corresponding author: a.ezeribe@sheffield.ac.uk.}
\author[c]{J.~-L.~Gauvreau,}
\author[d]{J.~L.~Harton,} 
\author[e]{R.~Lafler,}
\author[e]{E.~R.~Lee,} 
\author[e]{D.~Loomba,}
\author[c]{A.~Lumnah,}
\author[e]{E.~H.~Miller,}  
\author[b]{F.~Mouton,}  
\author[f]{A.~StJ.~Murphy,}  
\author[g]{S.~M.~Paling,}
\author[e]{N.~S.~Phan,}  
\author[b]{M.~Robinson,}
\author[b]{S.~W.~Sadler,} 
\author[b]{A.~Scarff,}
\author[d]{F.~G.~Schuckman~II,}  
\author[c]{D.~P.~Snowden-Ifft,} 
\author[b]{and N.~J.~C.~Spooner}

\affiliation[a]{Department of Physics, Wellesley College, 106 Central Street, Wellesley, MA 02481, USA}
\affiliation[b]{Department of Physics and Astronomy, University of Sheffield, S3 7RH, UK}
\affiliation[c]{Department of Physics, Occidental College, Los Angeles, CA 90041, USA }
\affiliation[d]{Department of Physics, Colorado State University, Fort Collins, CO 80523-1875, USA}
\affiliation[e]{Department of Physics and Astronomy, University of New Mexico, NM 87131, USA}
\affiliation[f]{School of Physics and Astronomy, University of Edinburgh, EH9 3FD, UK}
\affiliation[g]{STFC Boulby Underground Science Facility, Boulby Mine, Cleveland, TS13 4UZ, UK}

\abstract{Recent computational results suggest that directional dark matter detectors have potential to probe for WIMP dark matter particles below the neutrino floor.  The DRIFT-IId detector used in this work is a leading directional WIMP search time projection chamber detector. We report the first measurements of the detection of the directional nuclear recoils in a fully fiducialised low-pressure time projection chamber.  In this new operational mode, the distance between each event vertex and the readout plane is determined by the measurement of minority carriers produced by adding a small amount of oxygen to the nominal CS$_{2}$+CF$_{4}$ target gas mixture. The CS$_2$+CF$_4$+O$_2$ mixture has been shown to enable background-free operation at current sensitivities. Sulfur, fluorine, and carbon recoils were generated using neutrons emitted from a $^{252}$Cf source positioned at different locations around the detector.  Measurement of the relative energy loss along the recoil tracks allowed the track vector sense, or the so-called head-tail asymmetry parameter, to be deduced.  Results show that the previously reported observation of head-tail sensitivity in pure CS$_{2}$ is well retained after the addition of oxygen to the gas mixture.}


\begin{document}
\maketitle
\flushbottom

\section{Introduction} \label{sec:intro}
A central unsolved problem in physics is to unravel the nature of non-baryonic dark matter (DM), known to  constitute 24\si{\percent} of total energy density of the Universe \cite{bib1a}. Many possible DM candidates have been proposed.  Weakly Interacting Massive Particles (WIMPs), motivated in part by supersymmetric theories, are a popular candidate \cite{bib1ab, bib1aba}. There is currently a global effort towards detecting these putative WIMP dark matter particles, in particular, by direct dark matter search experiments that aim to observe the products of WIMP-nucleus elastic scattering \cite{bib1aba}. The rate of these events, and the recoil energy spectrum produced, is expected to modulate annually. This provides one possible route to a definitive signature related to our motion through the dark matter halo.    DAMA/LIBRA \cite{bib1b} have announced a cumulative positive result from such an analysis with over 9$\sigma$ statistical significance.     However, this result remains controversial \cite{bib1c, bib1d, bib1e, bib1f},  and the technique remains problematic since backgrounds may mimic a WIMP signal \cite{bib1d, bib2a, bib2b, bib2bc}.

A robust and potentially more powerful galactic signature is encoded in the direction of the WIMP-induced nuclear recoils.  The angular distribution of WIMP-induced recoils is highly anisotropic in galactic coordinates, with the average direction oriented away from our direction of travel towards the constellation Cygnus.  In contrast, the distribution of any terrestrial background would be expected to be uncorrelated with this direction.  In laboratory coordinates, due to the Earth's rotation, a sidereal oscillation in WIMP-induced recoil directions would be expected \cite{bib1a, bib2bc}.  By this means, directional experiments have sensitivity to DM cross-sections below the so-called neutrino floor imposed by solar and atmospheric neutrinos coherently scattering off target nuclei \cite{bib2c, bib2d}. For a recent review of the discovery reach of directional DM detectors, see Ref. \cite{bib2de}. 

An idealised detector sensitive to the recoil track axis only (not the vector direction), would need a few hundred events to confirm a galactic WIMP signal \cite{bib2e}. If, however, the recoil track vector is measured, then only tens of events are required \cite{bib2e,bib2f}.   An important component of the vector track information is the so-called head-tail information, which determines on which end of the track the collision vertex is located. Determination of head-tail information is of great importance to the development of directional WIMP search technology.  Although there has been recent progress in the use of nuclear emulsions in directional detection \cite{bib3a}, the leading experiments all use a gaseous Time Projection Chamber (TPC).  When operated at low pressure, typically 40-100 Torr, the expected recoils have a length of a few millimetres, long enough to be measured using a variety of existing charge sensitive readout technologies.   Importantly,  in principle this also allows measurement of any asymmetry in the ionization along the tracks and hence the head-tail of events.  Directional DM search experiments including DRIFT \cite{bib3b}, NEWAGE \cite{bib3c}, MIMAC \cite{bib3d}, DMTPC \cite{bib3e} and D$^{3}$ \cite{bib3f}  have been championing the use of this DM tracking method \cite{bib2de,bib3g}. These directional experiments are developing a joint initiative, called CYGNUS-TPC, to consider building a large modular directional experiment that would include head-tail sensitivity.

The DRIFT collaboration has already demonstrated head-tail sensitivity using pure CS$_{2}$ target gas  \cite{bib3b, bib4a}.  DRIFT chose to use this particular gas because it drifts signals as negative ions which results in much lower diffusion of tracks than can be achieved by conventional gases that drift electrons.  Since that work, a different gas mixture has been developed comprising a 30:10:1~Torr mix of CS$_{2}$:CF$_{4}$:O$_2$.  The relatively large ground-state nuclear spin of the fluorine nuclei provides a significant sensitivity to spin-dependent WIMP-proton interactions \cite{bib4ab}.  The addition of 1~Torr of oxygen gas to the existing CS$_2$+CF$_4$ mixture results in additional populations of faster minority charge carriers. Measurement of the relative times of arrival of these charge carriers at the $x-y$ charge readout plane can be used to determine the absolute position of the event vertex in the drift, that is $z$-axis, direction \cite{bib4b}.  This development of $z$-fiducialisation, adding to the existing $x$ and $y$ fiducialisation achieved by the MWPC (Multi-Wire Proportional Chamber) readout edge vetoes in DRIFT, has led to fully fiducialised operation, and a recently published background-free WIMP search -- a first for a directional DM experiment \cite{bib4c}.  The prime aim of the work reported here is to establish that the head-tail effect previously measured with pure CS$_2$ can still be detected in this new gas mode.

\section{DRIFT-IId detector and $z$-fiducialisation} \label{drift.detector}  
The DRIFT-IId directional dark matter detector comprises two back-to-back negative ion TPCs of 1~\si{\cubic \meter} total volume instrumented with MWPC readouts \cite{bib4e}. In this device, the two 50 \si{\centi\meter} long TPCs are separated by an aluminized-mylar thin-film central cathode biased at $-$31.9~\si{\kilo \volt}, creating a drift field of 580~\si{\volt \per \centi \meter} for anions drifting towards the two MWPCs \cite{bib4c, bib4d}. Each MWPC comprises parallel stainless steel anode wires of 20~\si{\micro \meter} diameter placed between and orthogonal to two grid wire planes made from parallel 100~\si{\micro \meter} diameter stainless steel wires at 1~\si{\centi \meter} anode-grid distance. The grid potential is set at $-$2.884~\si{\kilo\volt}, which is enough to strip off electrons from the drifting CS$_{2}$ anions and induce electron avalanches near the grounded anode wires. Each of the wire planes consist of 552 wires with 2~\si{\milli \meter} spacing. In the anode plane, 22 edge wires are grouped as guard wires, while the subsequent 82 wires are grouped as an anode veto. For the grid plane, 104 edge wires are grouped as a veto against side events. The remaining 448 wires each in the anode and grid planes are grouped down to 8 channels such that every 8$^{\text{th}}$ wire in each of the readout planes is read out by the same electronics channel.  Each of the 8 channels is then read out with a Cremat CR-111 charge sensitive pre-amplifier, followed by a Cremat CR-200-4~\si{\micro \second} shaping amplifier.  After signal amplification, the data are written to disk for analysis.
\begin{figure}[b!] 
\centering
\subfigure[High-z event at $z = 48.6$ \si{\centi \meter}, NIPs $= 2300$.]{
\includegraphics[width=0.46\linewidth,height=0.27\textheight]{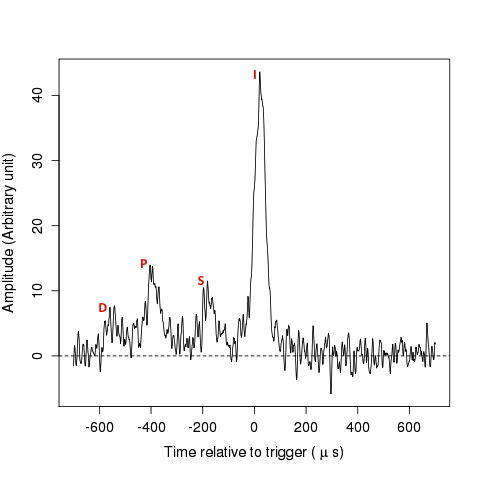} 
\label{fig:high-z}
}\hfil
\subfigure[Mid-z event at $z = 25.1$  \si{\centi \meter}, NIPs $= 1504$.]{
\includegraphics[width=0.46\linewidth,height=0.27\textheight]{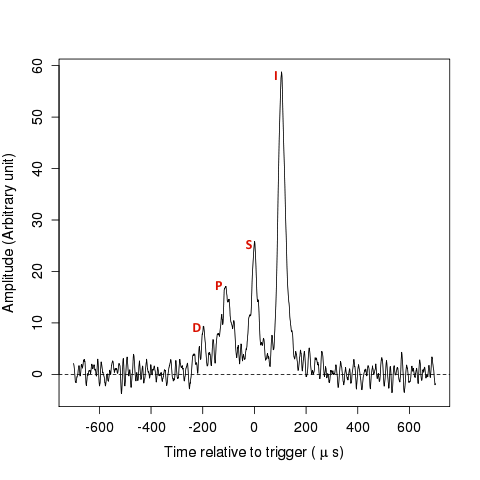} 
\label{fig:mid-z}
}
\subfigure[Low-z event at $z = 9.2$ \si{\centi \meter}, NIPs $= 1685$.]{
\includegraphics[width=0.46\linewidth,height=0.27\textheight]{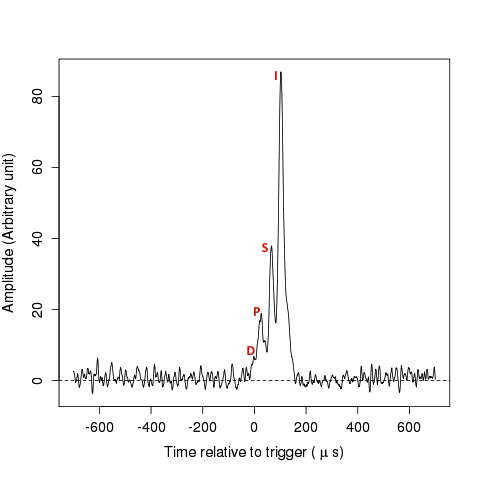} 
\label{fig:low-z1}
} \hfil
\subfigure[Low-z event at $z$ $=$ ?, NIPs $=$ 1851.]{
\includegraphics[width=0.46\linewidth,height=0.27\textheight]{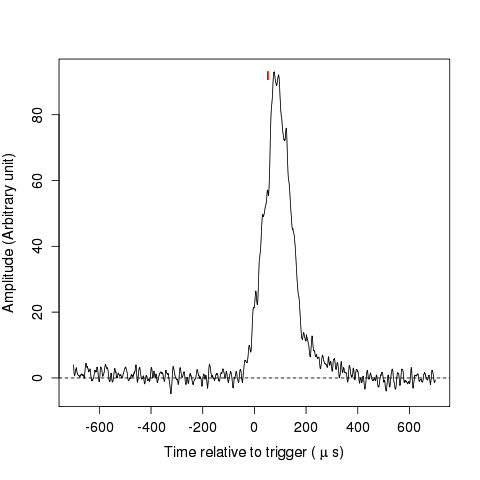} 
\label{fig:low-z2}
}
\caption{Signal pulses from neutron-induced recoils at different $z$-positions within the active volume of the detector. The S specie charge carriers can change to other species of anions during drift, which reduces amplitude of the S peak (see (a)). Distinct D, P, S, and I peaks are shown for a mid-$z$ event in (b).  Typical behaviour of these peaks as they merge in the low-$z$ region is shown in (c). The worst case in the very low-z region (closer to MWPC) is shown in (d). Events in the low-$z$ region were rejected during analysis since their minority carriers become concealed under the main charge cloud. NIPs here stands for number of ion pairs.}
\label{fig:peaks}
\end{figure}

The addition of 1~Torr of oxygen to the CS$_{2}$+CF$_{4}$ mixture results in several minority carriers in addition to the main charge cloud, each drifting at a slightly different speed in the DRIFT detector.   These minority carriers are believed to be due to different species of anions \cite{bib4b}. The three minority charge carriers and the main charge cloud are labeled as D, P, S, and I in Figure \ref{fig:peaks}, with I being the primary ionization, equivalent to the charge cloud observed in the absence of oxygen \cite{bib4b}. The minority carriers exhibit various characteristics with their behaviour depending mostly on $z$-position of the event as illustrated in Figures \ref{fig:high-z} to \ref{fig:low-z2}.  The four signal pulses shown in Figure \ref{fig:peaks} are from four different neutron induced nuclear recoils obtained from the anode wire that recorded the highest voltage after the respective events. 

The I peak contains about 50\si{\percent} of the total ionization products in each event while the remaining $\sim$50\si{\percent} is shared between the D, P, and S peaks \cite{bib4b}.  The separation in arrival times of the anions in these peaks at the $x-y$ charge readout plane allows for the absolute position of an event vertex along the $z$-axis to be determined, and hence the detector is fiducialised.  The absolute position of an event in the $z$-direction is given by: 
\begin{equation}
z =  \left( t_i - t_p  \right)  \left( \frac{ v_{i}v_{p} } {v_{p}-v_{i}} \right) \,,
\label{eq:z}
\end{equation}  
\nolinebreak
where $z$ is the distance between an event vertex and the readout plane, $t_i$ and $t_p$ are the arrival times of the anions in the I peak and P peak respectively, after drifting with corresponding velocities $v_i$ and $v_p$.  The events $z$ positions increase from 0 (MWPC) to 50~\si{\centi \meter} (central cathode) for each of the two detectors. 

For the purpose of this analysis, events in the detector were grouped into three bins according to their $z$-position: low-$z$ (0 $< z \leq $ 15 \si{\centi \meter}), mid-$z$ (15 \si{\centi \meter} $ < z \leq $ 35 \si{\centi \meter}) and high-$z$ (35 \si{\centi \meter} $< z \leq $ 50 \si{\centi \meter}).  Relative to the I peak, the amplitude of the S peak is known to decrease with increasing $z$ \cite{bib4b}.  This suppression of the S peak for the high-$z$ events can be seen in Figure \ref{fig:high-z}. In this work, many high-$z$ events did not have a clear S peak, and so the $z$-position of each event was calculated from timing of the I and P peaks.  For mid-$z$ events, clear and distinct separation between the main and minority peaks was observed as depicted in Figure \ref{fig:mid-z}.   In this region, the amount of charge in the peaks decreases from the I to the D peak, in that order.    In the low-$z$ region, closest to the MWPC, the total drift time of the anions is short, and so the temporal spacing between the peaks is small. For many of the low-$z$ events, the four peaks are not resolved in time and hence, were not precisely fiducialised. As a result, we omit the low-$z$ region from this analysis.  The implication is that data from 30\si{\percent} of the active volume of the detector were not included in this analysis.  New work to allow inclusion of these events is currently underway.

As described in \cite{bib4c, bib4d}, a population of nuclear recoil events near $z=50$ \si{\centi \meter} are produced by radioactive decays on the central cathode.  The effect of this contamination on the high-$z$ sample is negligible because (1) the intrinsic decay rate is low compared to the neutron interaction rate ($\sim$3 events/day vs. $\sim2.3\times10^4$ events/day), (2) about half of the total number of these radioactive decay events with $z>50$ \si{\centi\meter} were not included in the head-tail analysis.

In the work presented here, the high-$z$ and mid-$z$ bins were selected and used as a means to investigate the possible effect of diffusion on the head-tail asymmetry for events closer to the central cathode and middle of the detector respectively.

\section{Experimental set-up}\label{exposure} 
\begin{figure}[h!]
\centering
\includegraphics[width=.83\textwidth,height=0.47\textheight]{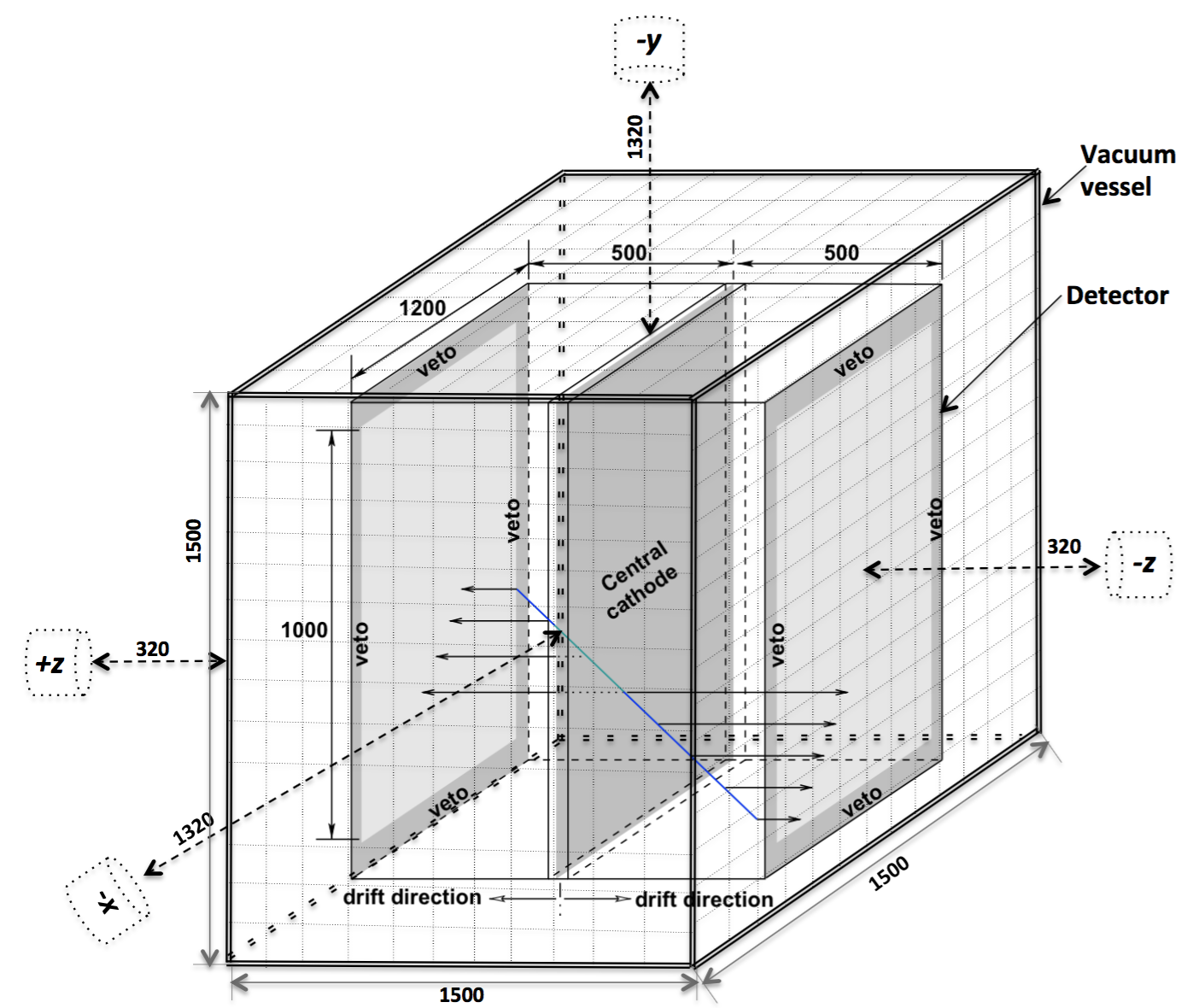}
\caption{DRIFT-IId detector showing different directions of neutron exposures, central cathode, veto, field direction and stainless steel vacuum vessel.  Positions of the $^{252}$Cf neutron source during the $+z$, $-x$, $-z$ and $-y$ runs are shown with small cylinders, which represent the lead canister used as the source container.  The dotted double-headed arrows show the separation distance between the neutron source and the vacuum vessel during the runs. There is a separation of 250~\si{\milli\meter}  (150~\si{\milli\meter}) between the detector and the vacuum vessel from the $+z$ and $-z$  ($-y$ and $-x$) directions.  Dimensions in \si{\milli \meter}.}
\label{fig:direction}
\end{figure}
The detector was exposed to neutrons from a $\sim$338 \si{\kilo \becquerel} $^{252}$Cf source from different directions for about seven days. These neutron exposures were performed without source collimation to minimise neutron back scattering. Illustration of the detector and details on the directions of the neutron exposures are depicted in Figure \ref{fig:direction}.  The hardware threshold was set to 15 \si{\milli \volt} throughout this experiment, resulting in a typical trigger rate of 5 \si{\hertz} and minimum detected recoil energy of $\sim$300 NIPs.  NIPs is the number of electron-ion pairs produced by recoiling nuclei.   Detector gain calibrations were performed once every 6 hours using X-rays from an internal 5.9 \si{\kilo \electronvolt} $^{55}$Fe source.  Throughout this experiment, an automated gas handling system established a continuous flow of 30:10:1 Torr CS$_2$:CF$_4$:O$_2$ at a rate of about 140 litres/hour (equivalent to about one detector volume change per day).

The source position was chosen such that the mean neutron direction pointed along the $z$-direction (perpendicular to the central cathode and MWPCs), $x$-direction (parallel to the grid wires), and $y$-direction (parallel to the anode wires).   In previous studies the DRIFT detector was found to be most sensitive to head-tail for events from the $z$-directions since microsecond sampling, coupled with the anion drift speed, results in a spatial sampling of about 60 \si{\micro \meter} along this direction. These are termed the optimal directions (see Ref. \cite{bib3b}). The detector was exposed from these optimal ($+z$ and $-z$) and anti-optimal ($-x$ and $-y$) directions.  In the exposure with the mean neutron direction (MND) along the $+z$ direction, the source was placed to the left of the MWPCs such that neutron induced recoils in the left detector were predominantly oriented towards the central cathode, while in the right detector neutron induced recoils were predominantly oriented away from the central cathode.   Inversely, in the exposure with the MND along the $-z$ direction, the source was placed to the right of the MWPCs such that neutron induced recoils in the right detector were predominantly oriented towards the central cathode, while in the left detector neutron induced recoils were mainly oriented away from the central cathode.   The runs where the source was centered on the central cathode from the $x$ and $y$-directions are known as $-x$ and $-y$ runs.  In each of these exposures, the source was placed at a distance of 32~\si{\centi\meter} (107~\si{\centi\meter}) and 132~\si{\centi \meter} (207~\si{\centi\meter}) away from the vessel (geometric center of the central cathode) for the optimal and anti-optimal directed neutron runs, respectively. These distances result in neutron interaction angular spread of about ~25\si{\degree} (14\si{\degree}) for the optimal (anti-optimal) runs. In these measurements, the source neutron-vessel separation distances for the optimal and anti-optimal runs were maximised subject to available space in the underground laboratory during the exposures. The exposure from the optimal directions was done to ascertain if the sensitivity of DRIFT-IId detector to signal head-tail (earlier reported in Ref. \cite{bib3b}) was affected by addition of oxygen that aided the full fiducialisation of the detector while the anti-optimal runs were expected to provide the required reference for null results.

\section{Data analysis}\label{analysis}
In this work, only the main charge clouds known as the I peaks were analysed to determine the head-tail effect. This is because the I peak retains most of an event charge compared to D, P, and S peaks.  The data analysis presented here was based on the method described in \cite{bib3b}. In the analysis, a nuclear recoil candidate must pass a series of initial cuts described in Ref. \cite{bib4d}. There are five main reasons to reject some events in this first stage. Events emanating from the central cathode and long alpha events may have hits (nuclear recoil signals that cross the analysis threshold) on both sides of the cathode - such events are rejected, as are events with hits on non-contiguous wires, which may be two independent events in the same time window.  An analysis threshold of 9 \si{\milli\volt} allowed for identification of the D, P, and S peaks since events are typically triggered by the I peak.  Apart from the I peak, at least any two of the D, P, and S peaks must pass the analysis threshold for an event to be analysed further as a nuclear recoil candidate.  Events with all eight wires in a group hit are cut to reject all contained alpha events with track length in the $x$-$y$ plane greater than 15~\si{\milli \meter}. This is because there are no contained alphas with $x$-$y$ range $<15$~\si{\milli\meter} in the DRIFT-IId detector design \cite{bib4c}. To reject other alphas and all other events emanating from outside the active volume of the detector, events with hits on the veto wires are rejected. Also, events with rise-times less than 3~\si{\micro \second} are rejected to remove noise (for instance sparks).

In the next stage of the analysis, the integral charge from each event in ADC units was converted to corresponding NIPs equivalent.  To do the NIPs conversions, a region of interest (ROI) was defined from $+700$~\si{\micro\second} to $-700$~\si{\micro\second} relative to the trigger time for all anode channels with signal above the analysis threshold. This ROI contains the D, P, S, and I peaks. The sum of the integrated charge values was converted to the NIPs equivalent value using the most recent $^{55}$Fe detector gain calibration constant and W-value of the gas mixture.  The W-value is the average amount of energy required to produce an electron-ion pair in a given gas mixture. 

The W-value for 30:10:1 Torr of CS$_{2}$:CF$_{4}$:O$_{2}$ gas is under study. But the addition of about 1\si{\percent} of O$_{2}$ to the conventional CS$_{2}$+CF$_{4}$ gas mixture is expected to change the W-value by only about 0.3 \si{\electronvolt} \cite{bib5a, bib5b, bib5c}. This is within the uncertainty of 25.2$\pm$0.6 \si{\electronvolt} measured by Ref. \cite{bib5d} for 30:10 Torr of CS$_{2}$:CF$_{4}$ mixture, adopted in this analysis.  The ionization energy depends on a property of the gas mixture known as the quenching or Lindhard factor (see Ref. \cite{bib5e}). This gas quenching factor measures the fraction of energy that goes into ionization for a given recoil.  For example, the quenching factor is 0.53 for a 50~\si{\kilo\electronvolt} fluorine recoil in a 30:10 Torr of CS$_2$:CF$_4$ gas mixture, and so a recoil of this energy will produce 1055 electron-ion pairs \cite{bib4d, bib5e}.  

\begin{figure}[h!]
\centering
\includegraphics[width=.62\textwidth,height=0.35\textheight]{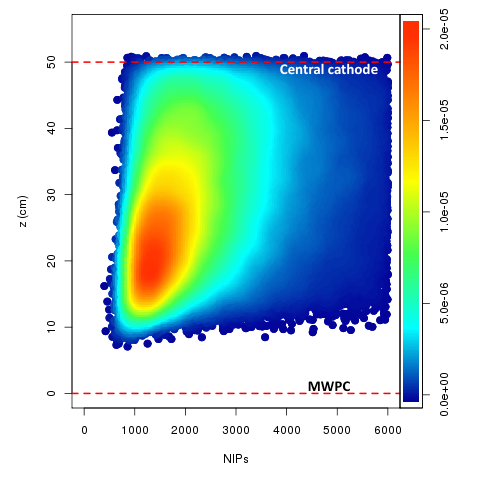}
\caption{Event distribution in the $z$-NIPs space obtained from neutron exposure using a $^{252}$Cf source placed in $+z$ position. This is before the $z$ cuts described in the text. The dashed red lines at 0 and 50 \si{\centi \meter} mark the locations of the readout plane and central cathode respectively. The colour scale represents the density of the events (colour online).}
\label{fig:zvse}
\end{figure}

Events with NIPs$\leq$6000 were considered further, no low-energy cut was applied.  See Figure \ref{fig:zvse} for the distribution of reconstructed neutron recoil events in the $z$-NIPs space before the $z$ cuts. The absolute $z$ position for each of the events was computed using Equation \ref{eq:z}. 
\begin{figure}[h!] 
\centering
\subfigure[Experimental data.]{
\includegraphics[width=0.485\linewidth,height=0.42\textheight]{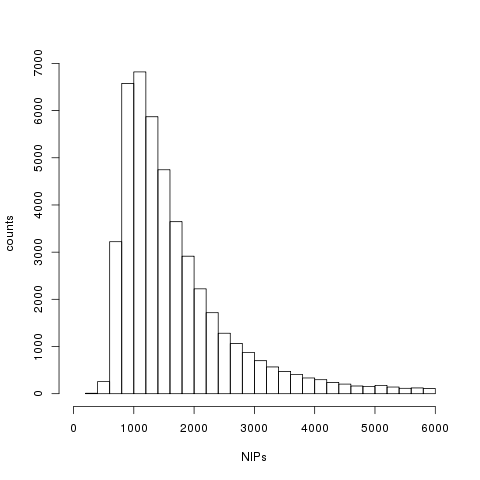} 
\label{fig:expspec}
}\hfil
\subfigure[GEANT4 results.]{
\includegraphics[width=0.485\linewidth,height=0.42\textheight]{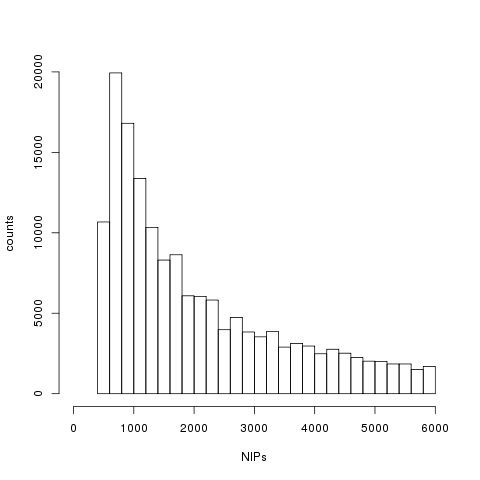} 
\label{fig:geant4spec}
}
\caption{Energy spectra of neutron-induced nuclear recoils with the source in the optimal direction. (a) Experimental data before the $z$ cuts (described in the text) from a $+z$ run. (b) Results of a GEANT4 Monte Carlo simulation.}
\label{fig:energyspectrum}
\end{figure}
There are no events in the left part of the panel in Figure \ref{fig:zvse} because charge signal pulses diffuse as they drift to the readout plane. For low-energy, high-$z$ events, diffusion broadens the pulse width and may reduce the maximum pulse amplitude below the data acquisition trigger. The trigger has since been modified to monitor the integral charge of the waveform, rather than the amplitude. The upgraded trigger should improve the efficiency for high-$z$ events at low-energy. In the lower part of the panel, closer to the readout plane, the reconstruction cannot identify at least two of the D, P, and S peaks because of the short drift time, and so these events do not pass nuclear recoil selection cuts. The energy distribution of these events shown in Figure \ref{fig:zvse} is compared to the expected energy spectrum from $^{12}$C, $^{19}$F and $^{32}$S recoil candidates generated with GEANT4 \cite{bib5ef} program in Figure \ref{fig:energyspectrum}.  It can be seen from these two spectra that the distribution of the experimental data and theoretical expectation from $^{12}$C, $^{19}$F, $^{32}$S candidates all reach maximum at about 1000 NIPs, each with an exponential fall-off at higher NIPs. Hence, it is likely that there are some fractions of $^{12}$C, $^{19}$F and $^{32}$S recoil candidates in this data set. However, the energy spectrum obtained from the GEANT4 simulation (see Figure \ref{fig:geant4spec}) has a slower high-energy fall-off with a steeper cut-off at low-energy when compared to the experimental data shown in Figure \ref{fig:expspec}.  These discrepancies may be related to unexplained systematics discussed in Section \ref{result}. %

Additionally, to quantify the relative amount of ionization in the minority peaks, a parameter called the peak ratio $Q$, defined as: 
\begin{equation}
Q =  \frac{ Q_{DPS}} {Q_T}  \,,
\label{eq:mp.ratio1}
\end{equation} 
\nolinebreak 
was also computed using the signal pulse on the anode wire that recorded the highest voltage in each event, where $Q_{DPS}$ is the integral charge in the D, P, and S peaks, and  $Q_T$ is the total integral charge in the D, P, S, and I peaks.  Events whose $Q$ values are greater than 0.30 and less than 0.65 were analysed further. This cut was found to be a powerful tool for discriminating nuclear recoils from radon progeny recoil events emanating from the MWPC charge readouts and some low-$z$ events whose minority charge carrier peaks are completely concealed under the main charge cloud.  These noise events are expected to have little or no minority charge carrier peaks, which result in small $Q_{DPS}$ values with $Q < 0.30$. The upper cut on $Q > 0.65$ is used to remove spark events with continuous waveform pulses resulting in large $Q_{DPS}$ values relative to expectations from nuclear recoil events.

\begin{figure}[h!] 
\centering
\includegraphics[width=.6\textwidth,height=0.4\textheight]{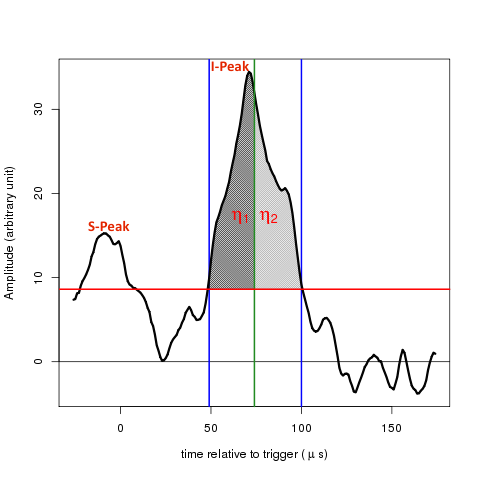}
\caption{Event waveform from a single wire showing the analysed I peak with  $\eta_{1}$ (integral charge in the first half of region of interest) shaded dark grey and $\eta_{2}$ (integral charge in the second half of the region of interest) shaded light grey. The two blue lines mark the region where the signal exceeds 25\si{\percent} of the peak amplitude. The green line is the mid-point time (half of signal duration above analysis threshold) between the first and second halves of the signal and the red line shows the 25\si{\percent} of the I peak used as the analysis threshold (colour online).}
\label{fig:nips.ratio}
\end{figure}

To investigate the presence of the head-tail effect in the nuclear recoil tracks, we use the charge distribution in the main charge cloud from every nuclear recoil event. This is because of the expectation that any available head-tail signature will cause a temporal asymmetry in the charge distribution along the nuclear recoil tracks \cite{bib3b,bib5f}. The asymmetry was measured by comparing the integral charge in the first and second halves of the I peak using the signal pulse in the anode wire with the highest voltage for that event. Specifically, the temporal extent of the I peak was defined as the region where the signal is larger than 25\si{\percent} of the peak amplitude (see Figure \ref{fig:nips.ratio}). 

This region was then split into two equal-length regions, and the integral charges $\eta_{1}$ and $\eta_{2}$ were computed for each of the events that passed the cuts. The vector direction of the track was then determined from the ratio:
\begin{equation}
\alpha \equiv  \frac{ \eta_{1} } {\eta_{2}} \,.
\label{eq:nips.ratio} 
\end{equation} 
\nolinebreak
Because the ionization density for nuclear recoils is larger at the start of the track than the end, a nuclear recoil whose velocity vector points towards the MWPC should have $\eta_1 < \eta_2$, and therefore $\alpha < 1$. Tracks with $\alpha > 1$ correspond to recoils that point toward the central cathode.  

The head-tail asymmetry parameter $\alpha$ was computed for the events in each TPC. Then, the mean value of $\alpha$ was computed separately for events in the left and right detectors: $\langle{\alpha}\rangle_L$ and $\langle{\alpha}\rangle_R$, respectively.  There is a temporal asymmetry introduced by the shaping electronics in the measured waveforms. To account for this effect, we compute $\Delta \alpha \equiv  \langle\alpha\rangle_{L} - \langle\alpha\rangle_{R}$ between the mean asymmetry parameters for the left and right detectors for each neutron exposure.  The expectation is that the largest head-tail effect should come from the $+z$ and $-z$ neutron source runs. In contrast, no head-tail effect is expected in the dataset from the anti-optimal $-x$ and $-y$ runs. 

To understand the head-tail effect further, the percentage difference of the head-tail asymmetry parameter was determined and termed $\delta$. This $\delta$ parameter, given by:
\begin{equation}
\delta =  100\frac{|\Delta\alpha|}   {\frac{1}{2}\left(\langle\alpha\rangle_{L} +  \langle\alpha\rangle_{R} \right) }   \,,
\label{eq:opd}
\end{equation}  \newline
\nolinebreak
is the ratio of $|\Delta\alpha|$ to the mean of $\langle\alpha\rangle_R$ and  $\langle\alpha\rangle_L$, expressed as a percent.  The $\delta$ parameter quantifies the measurability of the head-tail effect, including both the intrinsic head-tail signature and the sensitivity of the detector to that signature. The $\delta$ parameter was measured using events from the optimal and anti-optimal directions. The results obtained in this analysis are discussed in Section \ref{result}.

\section{Results and discussion} \label{result}
\begin{figure} [b!]  
\centering
\includegraphics[width=.55\textwidth,height=0.34\textheight]{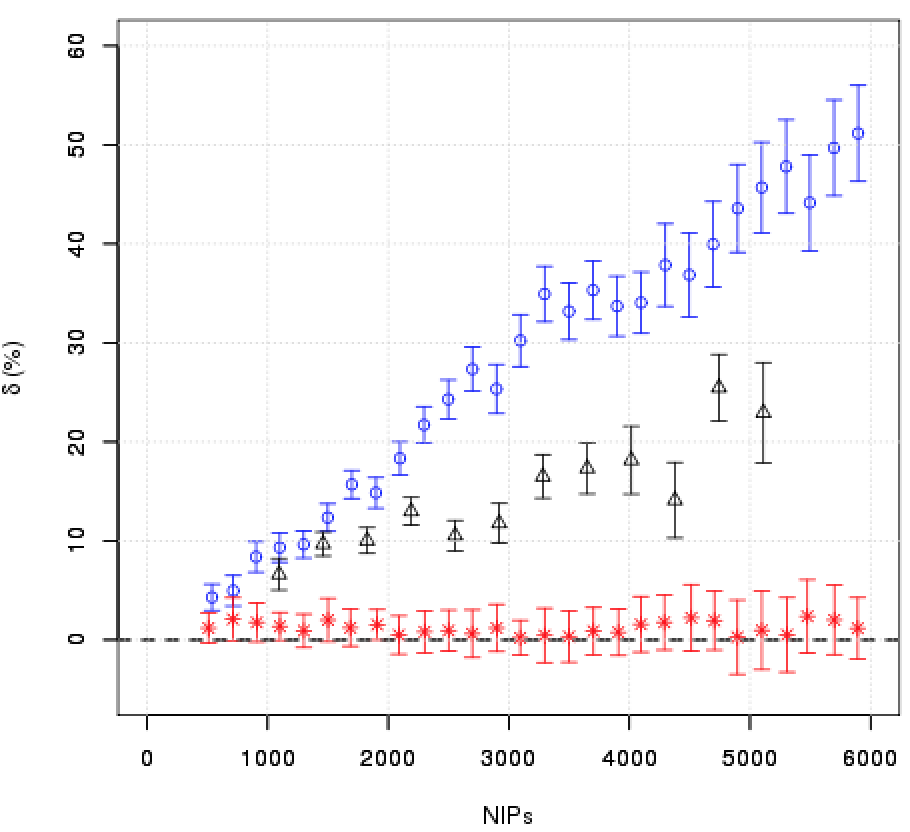}
\caption{Percentage difference $\delta$ parameter for different recoil energies represented by NIPs equivalent.  Results from this analysis (obtained with $^{12}$C, $^{19}$F and $^{32}$S tracks) are shown with blue circle points for the optimal runs and red asterisk points for the anti-optimal runs. The black triangle points are 2009 results obtained with only $^{32}$S tracks in pure CS$_2$ (see Ref. \cite{bib3b}). Colour online.}
\label{fig:per:diff}
\end{figure}
The dependence of $\delta$ on energy is shown in Figure  \ref{fig:per:diff} for both the optimal and the anti-optimal runs, using bins of width 200 NIPs.  It can be seen in that figure that the $\delta$ parameter measured in the anti-optimal runs is consistent with zero for all energies indicating that no head-tail signature was detected. This is expected because the mean directions of event tracks in these anti-optimal runs are the same for both TPCs, resulting in null $|\Delta\alpha|$ and $\delta$ parameters.  Conversely, events from the optimal runs, where the mean direction of recoil tracks is oriented along the drift direction of anions, show a $\delta$ parameter that increases with energy.  The $\delta$ parameter obtained from the optimal case is significant at 750 NIPs ($\sim$38 keV F-recoil energy equivalent). These results demonstrate the sensitivity of the detector to the head-tail directional signature and, furthermore, increased sensitivity with higher energy tracks.

In Figure \ref{fig:per:diff}, the $\delta$ parameter obtained in this new gas mode is compared to the result obtained in a previous measurement performed in 2009 using pure CS$_2$ gas \cite{bib3b}. It was shown in that study that event tracks used in the measurement were primarily from $^{32}$S recoils.  The lower analysis threshold used in this measurement yielded the observed head-tail sensitivity below the $\sim$1000 NIPs threshold obtained from the pure CS$_2$ data. Above 2000 NIPs, the measured values for $\delta$ in this work are approximately a factor of 2 larger than in the 2009 study.  

To explain this, we hypothesize that the observed increase in the $\delta$ parameter in Figure \ref{fig:per:diff} is due to the inclusion of $^{12}$C and $^{19}$F recoils in this data set which were not included in the 2009 study.  A longer recoil track is expected to yield a larger head-tail effect because it allows for a more distinct separation between the beginning and the end of the recoil track. For instance, a SRIM \cite{bib5g} calculation of the ranges for $^{12}$C, $^{19}$F, and $^{32}$S recoils that produce 1000 NIPs in 30:10:1 Torr of CS$_2$:CF$_4$:O$_2$ gives 1.70~\si{\milli \meter}, 1.51~\si{\milli \meter} and 0.98~\si{\milli \meter}, respectively.  However, the gas quenching for each of the $^{12}$C, $^{19}$F and $^{32}$S recoil candidates is different \cite{bib5f}. This gas quenching effect is expected to increase with mass of a given nucleus.  Hence, in the context of this experiment, $^{32}$S nuclear recoil tracks should suffer more energy loss per given recoil distance relative to $^{12}$C or $^{19}$F recoil candidates. Earlier studies in Ref. \cite{bib3c} reported the presence of head-tail effects in nuclear recoil tracks obtained with a CF$_4$ based TPC detector with potential $^{12}$C and $^{19}$F recoil candidates.

\begin{figure}[b!] 
\centering
\includegraphics[width=.65\textwidth,height=0.35\textheight]{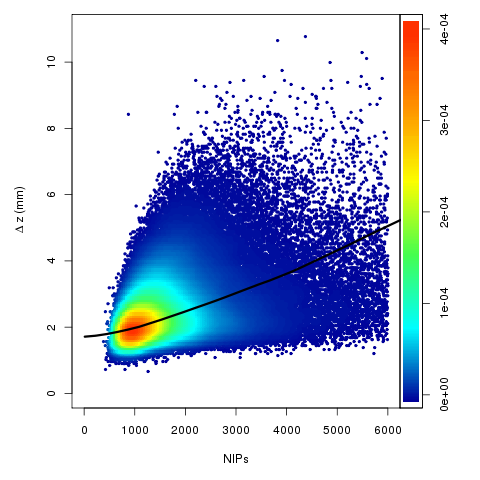}
\caption{Experimental data overlaid with the maximal $\Delta z$ predictions for $^{32}$S recoils in 30:10:1 Torr of CS$_2$:CF$_4$:O$_2$ gas mix obtained using SRIM.  The heat map is the experimental data while the black line is the maximum $\Delta z$ predictions for $^{32}$S recoils. The colour scale represents the density of the events (colour online).}
\label{fig:deltazdatasrim}
\end{figure}
To check for the presence of nuclear recoil candidates with track ranges that are greater than the expectations from only $^{32}$S recoils, the full width half maximum (FWHM) range $\Delta z = FWHM \times v_{i}$, was computed for each of the optimal ($+z$ and $-z$) events. This $\Delta z$ range was used because of its high resolution ($<$0.1~\si{\milli\meter}) \cite{bib5ga}. The results obtained from this analyses are shown in $\Delta z$-NIPs space in Figure \ref{fig:deltazdatasrim}. The SRIM range (in 3D) as a function of NIPs was computed for $^{32}$S, and the maximum expected $\Delta z$ parameter for $^{32}$S recoils was then determined by accounting for track diffusion, shaping electronics, and the differences in charge-carrier trajectories due to non-uniform electric field line distribution near the anode readout wires \cite{bib6a}.  The measurements show a significant number of events with $\Delta z$ in excess of the expectations for $^{32}$S, but consistent with presence of longer recoil tracks. However, the $\Delta z$-NIPs results did not unambiguously isolate each recoil specie candidates. Isolated recoil species will be a more convincing evidence to show the number of $^{12}$C and $^{19}$F recoil populations in this data. Ongoing studies are aimed toward tackling the ambiguity in separating constituent recoil species in a given data set.

The head-tail asymmetry parameter $\Delta\alpha$ obtained from the high-$z$ and mid-$z$ bins in the $+z$ and $-z$ runs are shown in Tables \ref{table:high:z} and \ref{table:mid:z}.   Significant head-tail effects are apparent in both the high-$z$ and mid-$z$ regions as well as the combined data in Table \ref{table:all}.  Naively, it is expected that a lower value of $\Delta\alpha$ will be observed in the high-$z$ region compared to the mid-$z$ region of the detector due to diffusion effects.  This is because diffusion can reorganise the actual charge distribution along a recoil track when it is exposed to longer drift distance which tends to conceal the head-tail effect. However, the opposite was observed because there are more high-energy (NIPs $>$ 1500) events in the high-$z$ region of the detector compared to the mid-$z$ region of the detector.  This is because the trigger efficiency on low-energy (NIPs $\leq$ 1500) events decreases for events closer to the central cathode due to accumulated diffusion effects along the drift direction. 

In fact, the fraction of high-energy events in the high-z region is about an order of magnitude greater than in the mid-z region of the detector. To investigate this effect, 19800 events were selected from the optimal high-$z$ and mid-$z$ regions using a fixed energy spectrum generated with $^{32}$S recoils in GEANT4. These selected events were analysed using the same procedure described above. The resultant $\Delta\alpha$ parameter of $0.098\pm0.005$ and $0.104\pm0.006$ were obtained from the high-$z$ and mid-$z$ regions, respectively. In this scenario, statistically consistent results were obtained from events in the high-$z$ and mid-$z$ regions. This result demonstrates that both diffusion and recoil energy affect the head-tail reconstruction. 

\begin{table}[h!] 
\centering
\tiny
\begin{adjustbox}{width= 15 cm}  
\tiny
\begin{tabular}{l*{7}{c}r@{${}={}$}r}
\hline \hline
MND    		& 		$n$	&	  live-time (days) 		&  		$\langle\alpha\rangle_L$ 		 & 		$\langle\alpha\rangle_R$ 		&  		$\Delta\alpha\pm\sigma_{\Delta\alpha}$  \\
\hline
$+z$       		 &     		19955	&	    1.909         &    	    1.294$\pm$0.003     	 &    	   1.146$\pm$0.004    	  &         0.148$\pm$0.005    \\ \\	 

$-z$      		 &    		 21054	&	    	3.713	 &   	    1.175$\pm$0.004     	  &         1.316$\pm$0.003  	  &    	      $-$0.141$\pm$0.005   \\ \\     
\hline \hline
\end{tabular} 
\end{adjustbox}
\caption{Head-tail parameter obtained from high-$z$ events in the $+z$, $-z$ runs. Only statistical uncertainties are quoted.  The $n$ is the number of events, other symbols used here are defined in the text.} 
\label{table:high:z}
\vspace*{0.25 cm}
\end{table}  
\begin{table}[h!] 
\tiny
\centering
\begin{adjustbox}{width= 15 cm}
\begin{tabular}{l*{7}{c}r@{${}={}$}r} 
\hline \hline
MND    		& 		$n$   	&	live-time (days)&  		$\langle\alpha\rangle_L$ 		 & 		$\langle\alpha\rangle_R$ 		&  		$\Delta\alpha\pm\sigma_{\Delta\alpha}$  	\\		
\hline
$+z$       		 &     		31267    	&	    1.909    &    	    1.359$\pm$0.003     	 &    	   1.256$\pm$0.004    	  &         0.103$\pm$0.005    \\ \\	   

$-z$      		 &    		49005	&	    3.713 &   	    1.303$\pm$0.003     	  &         1.377$\pm$0.002  	  &    	      $-$0.074$\pm$0.004   \\ \\    
\hline \hline
\end{tabular} 
\end{adjustbox}
\caption{Head-tail parameter obtained from mid-$z$ events in the $+z$, $-z$ runs. Only statistical uncertainties are quoted.  Symbols are defined in Table \ref{table:high:z}.} 
\label{table:mid:z}
\vspace*{0.25 cm}
\end{table}  
\begin{table}[h!] 
\small
\centering
\begin{adjustbox}{width= 15 cm}
\tiny
\begin{tabular}{l*{7}{c}r}
\hline \hline
MND    		& 		$n$  		&	 live-time (days)		&  		$\langle\alpha\rangle_L$ 		 & 		$\langle\alpha\rangle_R$ 		&  		$\Delta\alpha\pm\sigma_{\Delta\alpha}$  	\\
\hline
$+z$       		 &     		51222 	&	    1.909   	         &    	    1.336$\pm$0.002     	 &    	   1.203$\pm$0.003    	  &         0.133$\pm$0.003    	\\ \\

$-z$      		 &    		70059	&	    3.713	        &   	    1.258$\pm$0.003      	  &        1.360$\pm$0.002   	  &    	      $-$0.102$\pm$0.003   \\ \\ 

$-x$       		 &  		5094 	&	   0.915        &    	    1.408$\pm$0.009     	 &    	   1.411$\pm$0.009    	  &         $-$0.003$\pm$0.013    \\ \\

$-y$      		 &    		2873 	&	   0.561	        &   	    1.345$\pm$0.011      	  &        1.350$\pm$0.012   	  &    	      $-$0.005$\pm$0.016      \\ \\ 
\hline \hline
\end{tabular} 
\end{adjustbox}
\caption{Head-tail parameter obtained from combined high-$z$ and mid-$z$ events in the $+z$, $-z$, $-x$, and $-y$ runs. Only statistical uncertainties are quoted. Symbols are defined in Table \ref{table:high:z}. } 
\label{table:all}
\vspace*{0.25 cm}
\end{table}

The larger fraction of high-energy events in the high-$z$ region of the detector is expected to produce longer recoil tracks with distinct track head-tail separation and higher signal to noise ratio which yield higher values of the head-tail asymmetry parameter $\Delta\alpha$, compared to the mid-$z$ region.  Results obtained from all the events in the optimal ($+z$ and $-z$ data) and anti-optimal ($-x$ and $-y$ data) runs are presented in Table \ref{table:all}.  It can be seen from this table that a higher value of the head-tail parameter $\Delta\alpha$ was obtained from $+z$ and $-z$ directions when compared to  $-x$ and $-y$ runs.  This confirms that $+z$ and $-z$ are the optimal directions for head-tail detection in the DRIFT-IId detector design as reported in Ref. \cite{bib3b}.  Also as expected, $-x$ and $-y$ exposures returned null head-tail effect. It can be seen in Tables \ref{table:high:z} to \ref{table:all} that $\langle\alpha\rangle_R$ and $\langle\alpha\rangle_L$ for $+z$ and $-z$ neutrons, respectively, are not less than 1 as expected. As described in Section \ref{analysis}, this is due to the effect of the shaping electronics, which is accounted for in the computation of $\Delta\alpha$ parameter. Even with the effect of the shaping electronics, it can be seen that $\langle\alpha\rangle_L > \langle\alpha\rangle_R$ for the $+z$ run and $\langle\alpha\rangle_L < \langle\alpha\rangle_R$ for the $-z$ run, as expected.

Some cross-checks can be done on the results presented in Tables \ref{table:high:z}, \ref{table:mid:z} and \ref{table:all}. For each of the tables, the difference between the magnitude of the $\Delta\alpha$ parameter for the $+z$ and $-z$ runs should be zero within uncertainties. The differences are 0.007$\pm$0.007 (Table \ref{table:high:z}, high-$z$ region), 0.029$\pm$0.006 (Table \ref{table:mid:z}, mid-$z$ region) and 0.031$\pm$0.004 (Table \ref{table:all}, combined mid-$z$ and high-$z$ regions). These checks indicate agreement of the two measurements of $\Delta\alpha$ for the high-$z$ events, but also indicate the presence of a systematic effect for the two measurements of $\Delta\alpha$ for the mid-$z$ events of Table \ref{table:mid:z} and the combined data in Table \ref{table:all}.  Although, the observed head-tail asymmetry parameter remains statistically significant after allowing for this systematic effect. 

Additionally, two other alternative measures of the $\Delta\alpha$ parameter can be formulated by using only one side of the detector (left or right) with both $+z$ and $-z$ runs. This can be done by subtracting $\langle\alpha\rangle_L$ (for a $-z$ run) from $\langle\alpha\rangle_L$ (for a $+z$ run) and also by subtracting $\langle\alpha\rangle_R$ (for a $-z$ run) from $\langle\alpha\rangle_R$ (for a $+z$ run).  These two alternate calculations yielded 0.119$\pm$0.005 and -0.170$\pm$0.005 for the high-$z$ region of the detector. For the mid-$z$ events, the alternate measures were found to be 0.056$\pm$0.004 and -0.121$\pm$0.004, while for the combined runs the alternate measures are 0.078$\pm$0.004 and -0.157$\pm$0.004. The differences between the magnitude of each pair of these numbers obtained from the alternate measures are 0.051$\pm$0.007 (high-$z$), 0.065$\pm$0.006 (mid-$z$) and 0.079$\pm$0.006 (combined mid-$z$ and high-$z$).  The presence of these differences again indicate systematic effects, since the magnitude of the numbers obtained from the alternative measures are not equal within uncertainty. The differences reported above are estimates of the size of the systematic effects, which are larger than the statistical uncertainties reported in Tables \ref{table:high:z} - \ref{table:all}. 

After allowing for the systematic effects described above, the head-tail effect remains clearly present for the optimal runs. It should be noted that these systematic effects were not observed in \cite{bib3b}, however, the experimental conditions for this work were different in comparison to \cite{bib3b}. For mechanical reasons, the neutron source had to be placed closer to the detector for this work. In addition, for this work there existed an asymmetric distribution of materials surrounding the detector. GEANT4 simulations suggest that neutrons can be scattered from these materials, which likely contributed to the observed systematic effects. A detailed understanding of the systematics is beyond the scope of this paper, as the aim is to verify that the head-tail effect, observed in \cite{bib3b}, is still present while operating in the new fiducialised mode.

\section{Conclusion}\label{conclusion}
The data presented here indicate firstly that the DRIFT-IId detector is still sensitive to signal head-tail detection after addition of oxygen gas for fiducialisation. This is after the head-tail sensitivity was previously reported considering only $^{32}$S recoils in pure CS$_2$ target gas.  The longer recoil tracks due to $^{12}$C and $^{19}$F components in the new data were found to yield a more distinct separation between signal track head and tail. Studies done at different $z$ regions of the detector yielded higher values of the head-tail asymmetry parameter in the high-$z$ region compared to the mid-$z$ region.  This was found to be because the high-$z$ bin is dominated by high-energy events mostly because low-energy events in this region of the detector do not pass the analysis threshold due to larger diffusion effects.  

The $+z$ and $-z$ directions were confirmed as the optimal directional axes for the DRIFT-II detector while the $-x$ and $-y$ also produced null head-tail results, as expected. This observed head-tail effect from the optimal direction and null effects from the anti-optimal direction can be combined to track the modulation in signal directionality from WIMP events reaching the detector from the direction of the CYGNUS constellation. This method will be better utilised if the optimal axes of the detector are aligned along the earth north-south axis as proposed in previous studies.

These new results are encouraging for the possibility of developing a large volume TPC for directional dark matter detection where an appropriate readout technology, sensitive to head-tail effect can be combined with full detector fiducialisation.  Further work is underway to develop a more sophisticated analysis for extracting the concealed head-tail information due to peak pile-up in the low-$z$ region which is expected to improve the head-tail sensitivity further. Studies of alternative head-tail analyses are also underway to improve the overall directional sensitivity.

\acknowledgments
We acknowledge support from US National Science Foundation (NSF).    J.B.R.B. acknowledges the support of the Alfred P. Sloan Foundation (BR2012-011), and the Research Corporation for Science Advancement (Award Number: 23325).    A.C.E. is grateful for support from federal government of Nigeria through 2011/2012 (merged) AST\&D TETFund grant.    D.L. acknowledges support from NSF grant numbers 1407773 and 1506329.    The DRIFT collaboration would like to thank the management of Cleveland Potash Ltd for their continued support.

\end{document}